\documentclass[preprint]{elsarticle}

\usepackage[T1]{fontenc}

\DeclareRobustCommand{\VAN}[3]{#2}
\let\VANthebibliography\thebibliography
\def\thebibliography{\DeclareRobustCommand{\VAN}[3]{##3}\VANthebibliography}
\usepackage{amssymb}
\usepackage{color}
\newcommand{\etal}{\mbox{\rm{et al.}~~}}
\newcommand{\Kappa}{K}
\newcommand{\kms}{\mbox{km\,s$^{-1}$}}

\usepackage{graphicx}	
\usepackage{amsmath}	

\title{Do quasars form from primordial black holes ?}

\author[Jeremy Mould]{
Jeremy Mould,$^{1,2}$\thanks{E-mail: jmould@swin.edu.au}
and Adam Batten$^{1}$
\\
$^{1}$Swinburne University, PO Box 218, Hawthorn Vic 3122, Australia\\
$^{2}$ARC Centre of Excellence for Dark Matter Particle Physics\\
}

\date{Accepted XXX. Received YYY; in original form ZZZ}

\begin{document}
\label{firstpage}
\maketitle

	\section*{Abstract}
	We explore the consequences of a novel but increasingly well-supported hypothesis that supermassive black holes may have formed from primordial black holes (PBH) formed prior to, and rapidly growing in, the radiation-dominated universe.
	We show that this hypothesis  and fuelling by accretion of dwarf galaxies, made up of predominantly subsolar mass PBH and accompanying gas, can predict the luminosity of quasars and their distribution in luminosity. With reasonable values of the parameters introduced, these predictions are borne out by observations.
The model predicts density evolution in accordance with observations. 
If the same galaxy interaction rate creates quasars and radio galaxies,  whose primordial black hole nuclei seem somewhat less massive, their relative number densities reflect relative lifetimes in these states.

Primordial black holes(1292) - Cosmology(343) - Quasars(1319) - Supermassive black holes(1663) - Radio galaxies(1343)	





\section{Introduction}
Two breakthroughs, their discovery (Schmidt 1963) and identification of their black hole engines   (Lynden-Bell \& Rees 1971), marked the arrival of a new era of higher redshift astronomy.  Their
 physical understanding has been one of astronomy's greatest challenges.
This is because the QSOs' efficiency in energy production causes them to employ every conceivable channel for energy transfer right across the electromagnetic spectrum, plus radiation driven mass loss.
A review for $Nature$ by Antonucci (2013) claimed "Quasars still defy
explanation", and, while much knowledge has been added, that statement
remains true today.

A different way of looking at the physics is to consider their fuelling.
This has worked well with stellar evolution, but, as in 
stellar evolution, it tells us about
the numbers of objects at different stages, not, for example, details of their spectral energy distribution,
which is a product of much more involved physics,     see e.g.	 Porras-Valverde \etal ( 2025).
Here we consider their possible formation as primordial black holes (PBH $\S$2) 
(Bicknell \& Henricksen 1979; Volonteri, Habouzit \& Colpi 2021; 
Davies, Miller \&  Bellovary, 2011, Lupi \etal 2014;
Belotsky \etal 2019;
Ziparo, Gallerani \& Ferrara 2025; Cai \etal 2024;  Dayal \& Maiolino 2025;
Sobrinho \& Augusto 2024; Li, Huang \& Piao 2025, 
Zhang \etal 2025;  Hooper, Ireland, \& Krnjaic 2025; Liu \& Bromm 2025; 
Prole \& Regan 2025 
) during the radiation dominated era (RE) of the universe  (Mould 2025). 
Parameterizing variables that we don't know, we find we can successfully 
conform the model to 
their average bolometric luminosity ($\S$2) and  the luminosity function in $\S$3.
 In $\S$4 we draw attention to the imprint of inflation on the
origin of the seeds of SMBH, and conclude that the model can be further tested
with galaxy formation simulations and large volume surveys for QSOs with
JWST, and also the forthcoming $Roman$ space telescope and other observations.
\section{Luminosity density evolution}
\subsection{Initial  conditions}
According to the Friedmann equation the density during the RE is given by 1/$\rho$ = 32$\pi Gt^2$/3 or r = ct/8, where t is the age of the universe and r is the Schwarzschild radius of the primordial black hole that would form. There is an extensive literature on the seeds of PBHs originating during inflation. 
Reviews are given by \"Ogsoy \& Tasinato (2023) and Harada (2024). 
This RE radius equation  (which neglects the density excess that is
required to actually form a PBH
) allows  PBH as large as 
10~million M$_\odot$  to form at t = 400 seconds
at the end of primordial nucleosynthesis.  
In addition to supposing SMBH are PBH, we assume that a fraction f of the dark matter in the Universe is in the
form of PBH of mass 10$^{-8}$ M$_\odot$ (denoted m8) with an initial mass function (IMF)
that places equal mass in  each decade\footnote{A local IMF, not one that stretches very many orders of magnitude.
PBH formation may be associated with well separated phase transitions (Kuhnel, Rampf \& Sandstad 2016).} of mass (Mould 2025), n $\sim$ m$^{-1}$.
Given the density of dark matter from the Planck collaboration (2020), we can
write for the number density of m8
$$n(m8)=\frac{1.88 \times 10^{-29}}{2 \times 10^{25}} f \Omega_c h^2 (1+z)^3~~~~~~~{\rm gm/cc}~~\eqno(1)$$
at redshift z, where $\Omega_ch^2$ = 0.12. 
The choice of m8 is somewhat influenced by the end of constraints from the MACHO, OGLE and Eros microlensing 
projects (Alcock \etal 1998, Soszynski \etal 2024 and Moniez \etal 2017 ), but a different choice could equally well be normalized
with equation (1).

	%
\subsection{CMB distortions}
Limits have been placed on SMBH existing at the epoch of last scattering of the cosmic microwave background (CMB) 
by Nakama, Carr \& Silk (2018). A high  density of 
 the 
SMBH will lead to a $\mu$-distortion, an additive term to h$\nu$/kT in the Planck exponential.
The FIRAS limit $\mu~<$ 9 $\times$ 10$^{-5}$ (Fixsen \etal 1996) requires f$~^\prime~\lesssim$ 10$^{-8}$,
where f$~^\prime$ is the fraction of the dark matter in M7 (i.e. 10$^7$ M$_\odot$) PBHs.
We normalize our simplifying assumption that  all the dark matter is 10$^{-8}$ M$_\odot$  PBHs to f $\Omega_{matter}$ in the Planck cosmology. 
This automatically satisfies the CMB limit on f$^{~\prime}$.
Frampton (2018) sets an upper
limit on SMBH mass of 10$^{12}$ M$_\odot$ from this cause.
More recent papers (Deng 2021, Zegeye, Inomata \& Hu  2022, Yang \etal 2025)
anticipate a future limit on SMBH in the range (10$^6$, 10$^{15}$) M$_\odot%
$ of $\mu~<$ 10$^{-8}$. Wang, Huang \& Piao (2025) note the potential of
non-Gaussian fluctuations to alleviate this constraint.
\subsection{PBH IMF}
K\"uhnel, Rampf \& Sandstad (2016)
studied the formation of PBH in several models of the early Universe. 
PBH production has
generally been approximated to yield black holes of horizon
size at the time of formation. However, it was found and
subsequently confirmed in several thorough numerical works
that the PBHs are formed through critical
collapse. The critical collapse leads to a spectrum
for the formation of PBHs at any time over a wide range of masses, including 10$^{-11}$ to 10$^5$ M$_\odot$.
Here we adopt a minimalist assumption that equal masses exist in decades of mass (n $\sim$ m$^{-1}$) around particular peaks,
exemplified by K\"uhnel, Rampf \& Sandstadt's figure 4, and, for present purposes, we concentrate (1)
on subsolar masses making up the dominant DM in dwarf galaxies, and (2) SMBH.
The mass cut-offs in the subsolar regime can be adopted from microlensing observations: the "asteroid window" ranges from 10$^{-11}$ to 10$^{-7}$ M$_\odot$ 
(Niikura \etal 2019; Sugiyama, Takada, Yasuda \& Tominaga 2026).
We return to this subject in the next section, and to the supermassive range in $\S$2.7.
\subsection{Fuel burning}
Our model has a density of dark matter and needs a density of QSOs and
an encounter velocity to determine the collision rate.
We define a baseline QSO luminosity L46, where in one second m8 c$^2$ = 1.8 L46 ergs/s, and further choose 10$^7$ M$_\odot$ to represent the SMBH mass.
Whether the SMBH is part of a galaxy halo or not is not specified.
Substituting for  m8 from equation (1), the collision rate of SMBH and m8 PBH with a gravitational boost b, 
b $>$ 1, and $\sigma$ as a circular geometric cross section is$$ 
b~n_{M7}~ n_{m8}~ \sigma~ v = 
0.113 \times 10^{-54} f (1+z)^3 b v \pi (GM/c^2)^2 n_{M7} ~~~~{\rm cm^{-3}~s^{-1}}~~~\eqno(2)   $$
This collision rate is for SMBH and 10$^{-8}$ M$_\odot$ PBH, but the identification with dwarf
galaxies is made for dwarf galaxy sized clumps of dark matter, and the gaps in time 
between them correspond to QSO inactive periods. This is a standard
collision rate formula (Clayton 1968). 
Equation (2) is therefore
appropriate for SMBH activated by dwarf accretion.     
Orbital decay of the dwarf may be slow (Conselice \etal 2020).
we consider this part of the inactive period. Supposing $\kappa$100\% conversion of mass (all of the dwarf is consumable\footnote{Maximum values of $\kappa$ between 0.1 and 0.4 are considered possible (Blandford \& Begelman 1999; Kovacs \etal 2011)}, baryons and dark matter), the luminosity density of  quasars is 
$$\rho_L =  0.457 \times 10^{-32} n_{M7}
~ b~v~ \kappa~f~\pi   (1+z)^3 M7^2~~~~
{\rm ergs~ cm^{-3}~ s^{-1}}, \eqno(3)$$   
However, the cross section and mass increase with the accreted non-burned mass. 
The rate multiplier  is $\Kappa$ = (1+$\kappa(1-\kappa))\kappa ~>~ \kappa $ for all $\kappa~<~1$ 
for luminosity density evolution.
For example, for $\kappa$ = 0.4, $\Kappa$ = 0.5. 
With v$_{200}$ the velocity  in 200 km/s units (Peebles 1976, the cosmic virial theorem) 
the inverse of the volume occupied by an L46 QSO is $$
9.14~ 3.08^3 f \pi  \Kappa~ b~ v_{200}  ~(1+z_{obs})^3 n_{M7}  M7^2~~~~~{\rm Mpc}^{-3} 
\eqno(4)$$ 
and a range of 0.5 dex in M7  would give a range of 1 dex in luminosity.
Shen \etal (2020) find  this to be $\log\phi$ = 
 --4.7 cMpc$^{-3}$/dex for L = L$_{46}$ at z = 1, 
 compared with --3.8 for f = K = b = 1. Choosing to normalize at z = 2 would change this number to --4.8, compared to --5.0 from Shen \etal.
\subsection{Parameter constraints}
With~~b $>$ 1, as gravity is an attracting force, 
and z$_{obs}$ = 1 because it is very well determined observationally,  equation (4) becomes 

$$\log  n_{M7} \lesssim -4.7 -2.43 -\log f\Kappa~ v_{200}  -2 \log M7~~~~~~\eqno(5) $$
for b $>$ 1. 
Gravitational focussing can inform the value of b, and suggests b~=~1~+~(v$_{esc}/\sigma)^2$.
For a SMBH this is GM/R/$\sigma^2$, where R is the radius of the accretion disk,.
$\approx$ 1 + M/10$^5/\sigma_{30}^2~\approx $ 1, where $\sigma_30$ is the surrounding velocity dispersion in 30 \kms unitsi and M is in solar units..
For two compact objects, bound capture requires some form of dissipation, and this reduces the $\sigma$ term,
but not in a ready physically quantifiable way. If dissipation took 50\% of the energy per unit mass, $\sigma^2$,
for an M7 SMBH, b = 1 + 200 (30/200)$^2$ = 5.5 for $\sigma$~=~200 \kms.

A value of b and v$_{200}$ can be obtained from simulations by comparing the actual interaction rate of galaxies with the reaction rate with geometric cross sections. The halo mass function would also be involved. In the nearby SMBH sample of Winkel \etal (2025) $<M7>~\approx$ 2, but the 4 objects mentioned
in the database NED\footnote{ned.ipac.caltech.edu, 3C273, PG0026+129, Mrk1501, PG1211+143 
} as QSOs, as opposed to Seyfert galaxies, have a mean M7 of 23 and a dispersion of 0.5 dex. 
At present microlensing experiments provide the best way of estimating f.
Our setting the PBH compaction function to one instead of the current best estimate,
one fifth, has increased the predicted QSO luminosity density by a factor of
1.4$^6$. Correction is degenerate with the parameters f, $\Kappa$ and b.
\subsection{Physical Context of the accretion and fuelling}
QSOs show a characteristic big blue bump 
feature in the UV (Mandal, Woo \& Wang 2025), 
with the total luminosity at $\lambda~>$ 1$\mu$m typically 20\%-40\% of that at $\lambda~>$ 1$\mu$m. 
In this model the emission in the spectral range $\sim$ 10--300 nm, the "big blue bump, is dominated by 10,000-100,000 K thermal emission from an accretion disk. The emission between 2 microns and 1 mm, the "infrared bump," is made up of reradiation from dust in a distorted disk extending from 0.1 pc to more than 1 kpc (Sanders, Phinney, Neugebauer, Soifer \& Matthews 1989). 
The UV radiation  emanates from a concentrated
region surrounding the SMBH, which forms a geometrically thin, optically thick accretion disk, which
emits thermal radiation in the X-ray to optical wavelength range with a peak in the UV 
(Shakura \& Sunyaev 1973).
This is a vital component of our model, and the crux of the accretion scenario, but we deal here only with the bolometric output.

\subsection{Intermediate mass black holes}
In the paradigm in which there are only stellar mass black holes and SMBH, 
intermediate mass black holes (IMBH) are a conundrum. They arise naturally from the present PBH hypothesis, however. A second issue with IMBH is observational 
(Greene, Strader, \& Ho 2020).
The stellar mass, SMBH mass relation becomes sparse at 10$^5$ M$_\odot$. This, and the uncertainty around globular cluster nuclei, is most likely an observational selection effect,   whose cure is improved adaptive optics, to get close to the nucleus.
 The mass determined for a nuclear black hole is v$^2$r/G, where v is the velocity width of an emission line and r is the distance from the nucleus
of the ring of emitting gas. The smaller r, the further into IMBH territory the data require the mass derived to be.
IMBHs remain a puzzle in astrophysics because M $>$ 10$^4$ M$_\odot$ black holes cannot be stellar remnants (Davis \etal 2024), unless perhaps in galaxies earlier than those we can observe, in Population III. Formation as PBHs is, however, possible in principle.
If IMBH are confirmed in galactic or globular cluster nuclei, for example, 
it would be a further demonstration of a role for PBHs, in addition to the involvement in QSO formation we are discussing here.
\section{The quasar luminosity function}
Early work on the luminosity function of quasars (QLF) showed rapid evolution
with (1+z) (Schmidt 1968). Surveys have advanced the field into the epoch of reionization (Fan 2010).
The purpose of our examining the QLF is to see if any of the model assumptions with respect to PBH
lead to sharp disagreement with observations. There are a number of other QLF models that inform us about
the duty cycle of QSOs (Ren \& Trenti 2021), different SMBH populations,  downsizing and Eddington ratios
(Zhang \etal 2024). The former model is semi-empirical, and five parameters are fitted to the data, a
Schechter function arising from the halo mass function from which host galaxies are drawn,
the duty cycle being the free parameter. In our fuel burning model 
QSO activity would last 1.5 $\times$ 10$^8$ years, based on average luminosity and 10$^8$ M$_\odot$ of dwarf fuel.

The previous section gave us the expected number density of quasars accreting in this way. We combine this with the PBH IMF 
for the SMBH PBHs from Mould (2025),  log n + log M = constant, and    with a little algebra\footnote{ From equation (3)  
n(L) = $\rho_L/L ~\sim~ M^{-2}$ for L $~\sim~ M^4$ and
dlogn = -2 dlogM 
and, integrating over the IMF,
log n = 2 $\int$ 1/m dlogm.},
 we obtain Figure 1. For this PBH IMF, the form of the curve  from the previous section is obtained by integrating over it to get

 $$\log n = -4.7 + \frac{4x'}{\rm ln 100}~ (~10^{(42-x)/4} - 1~) \eqno(6)$$
where x = log L, x$^\prime$ = x - 46 and L42 is the luminosity of the minimum mass PBH progenitor, accounting for (42~-~x) in the exponent. This luminosity function is a Schechter function (power law with exponential cut off, Schechter 1976). 
It would be useful to find out if this is a better or worse fit to actual data than a double power law\footnote{A double power law can be formed from eqn (6) by bisecting it at L46.}. There is one less degree of freedom in the single power law and (x$_0$ = 42,  y$_0$ = -4.7) are fully independent, the y coordinate
being the y-axis of Figure 1.
The value of the minimum luminosity QSO, four orders of magnitude below L46, points directly to the post RE mass of its SMBH progenitor which would be a
 10$^6$ M$_\odot$ PBH. 'Minimum luminosity' should be regarded as a fitting parameter, rather than a physical constraint.
Furthermore, Mould (2025) shows that the cut-on mass of the IMF is a sharp marker of the cosmic temperature. 
The excellent fit\footnote{Shen \etal (2020) fits to the data at z = 1. in black, model in green} of the log M + log n = constant IMF is an indicator
that density inhomogeneities  with n $\sim$ M$^{-1}$ created during inflation were responsible
for the PBH IMF. 

\begin{figure}
\vspace{-1cm}
\includegraphics[angle=-0,width=.8\textwidth]{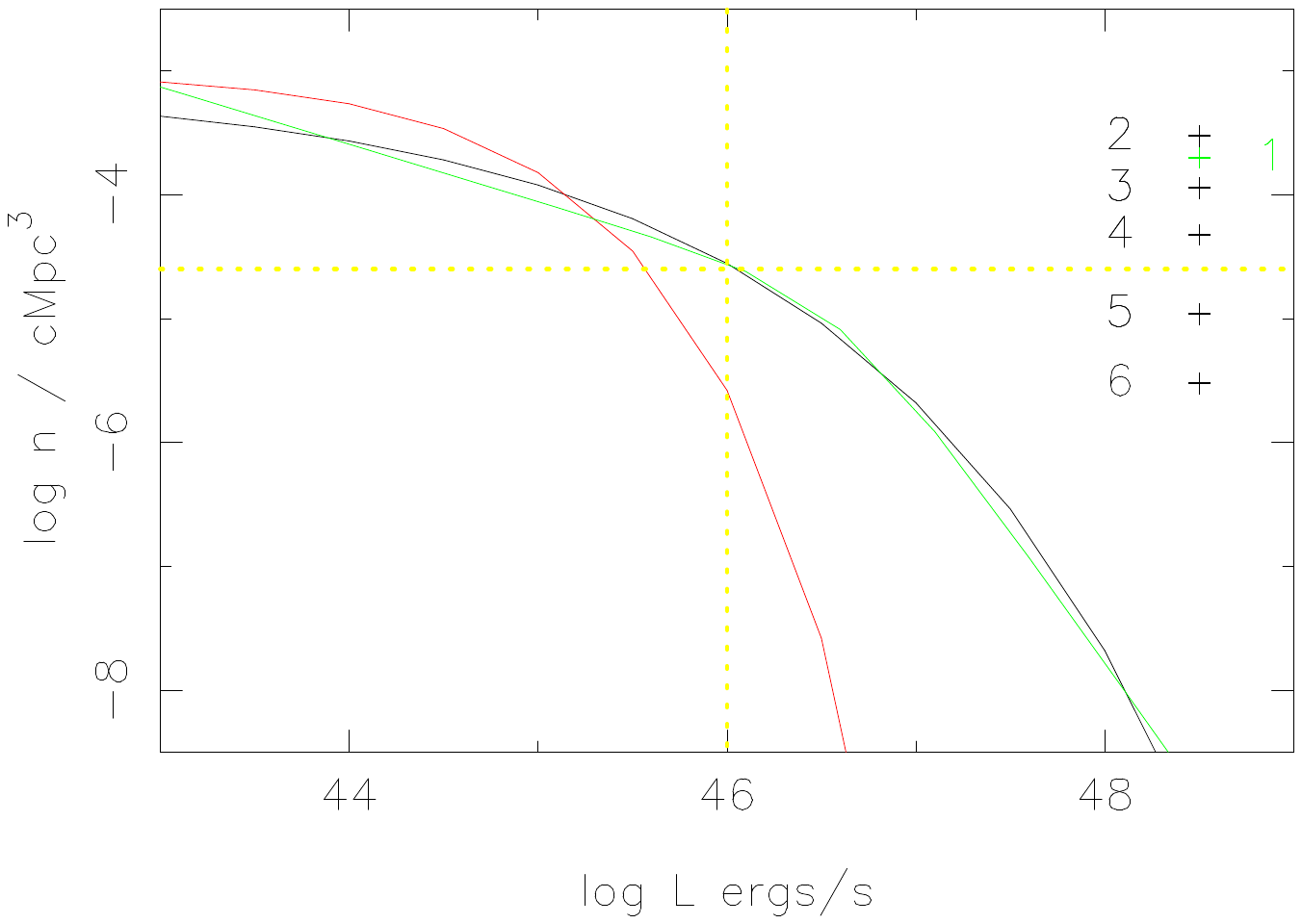}
	\caption{Prediction of the  QSO bolometric luminosity function in $\S$3 in black and the 
	double power law fit to the data at z = 1
	by Shen \etal (2020) in green. The model is normalised at L46 for appropriate values of the parameters in equation (5). In red is the prediction for PBH seeded
	in the RE  (Musco 2019), rather than during inflation,
which does not fit the data. The crosses on the right mark the density evolution from z = 1 to 6 at constant L46 luminosity.}
\end{figure}
\subsection{Parameter sensitivity}
The normalization of Figure 1 is equivalent to setting f = 0.125 in equation (4). This is a plausible value for the fraction of the DM that is PBH,
given the microlensing constraints. But f is degenerate with K and b in equation (4), and they have the same effect on Figure 1: moving the black curve
up and down. If b = 5, then f = 0.025 would restore the fit.

In the paradigm of $\S$2 it is interesting to ask why  there is a QSO luminosity function (LF) rather than a delta function at L46. And is the mean QSO luminosity a standard candle~?
Two prime candidates for sources of a distribution function are the m8 and M7 parameters in the model. 
The PBH IMF creates spreads in both
parameters. If only these parameters have a range, density evolution alone would be expected. 
A diverse formation history of dwarf galaxies across the relevant redshift range would lead to luminosity evolution as well.
For $\Kappa $ = 1 a fuelling rate of 10$^8 ~M_\odot $ per 10$^8 $ years (a free fall time) produces 6 $\times$ 10$^{46}$ ergs/s. The astrophysics that modulates this luminosity on shorter timescales is certainly complex, and a subject in itself. 
But it is the economics of supply that was considered in $\S$2. If the overall efficiency of consumption of galaxies by quasars were, say, 0.01\%, this could  be accommodated in equation (5) by an order of magnitude change in M7, the principle would still hold, and the LF would remain as shown.
\subsection{Luminosity and density evolution}
Equation (4) gives the dependence of quasar luminosity density on the
redshift of the observer, $\Delta\log n/\Delta(1+z)$ = --3. Figure 2 shows density evolution versus the scale factor. A slope of
-3 is expected from the model, and this is close to the data. Luminosity
evolution would require a variation of one or more of the parameters
with redshift to remain within the framework of the model. Both m8 (the fuel) and M7 (the
consumer) 
are candidates. 
Since the mass of the SMBH or its mean value was laid down at inflation, it is effectively ruled out
as a variable across the observable redshift range. The supply of dwarf
galaxies as fuel, however, is plausible as the responsible factor.
Observations can test this paradigm, both of galaxy number density versus
mass and redshift and of QSOs at JWST redshifts (Labb\'e \etal 2025) to extend Figure~2. 
The large error bar on the z $\sim$ 10 point in Figure 2
allows the pure density evolution of the model to fit the data. But
a future square degree sample of 100 or more QSOs, confirmed with spectra, would shrink the errors bars to $\pm$0.05 dex.
Matthee   
 \etal (2024)
 measure number densities of 10$^{-5}$ cMpc$^{-3}$.  
Kocevski \etal (2024) find that "little red dots" are 100--1000 more numerous than bright quasars at z $\approx$ 5--7, but their number density is only 4--10 times higher than X-ray and UV selected AGN. 

\begin{figure}
\vspace{-1.15cm}
\includegraphics[angle=-0,width=.9\textwidth]{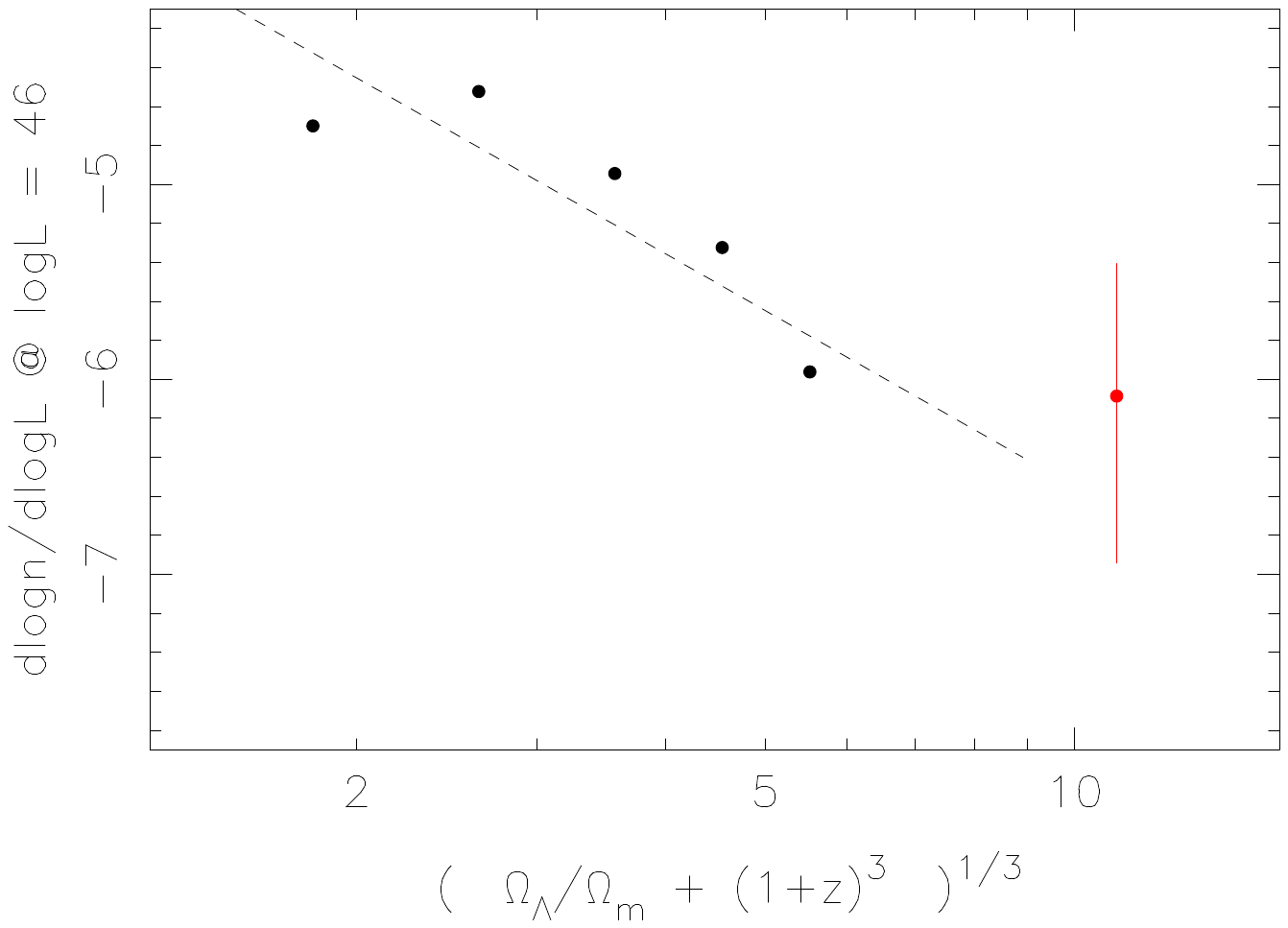}
\caption{The variation of number density at L46 with redshift in the
	data of Shen \etal (2020). The x-axis $\sim$ (1+z), but bridges
	the matter dominated and the dark energy dominated eras in a $\Lambda$CDM universe. 
The density evolution model of $\S$2 predicts a slope like the dashed line.	The red dot with Poisson error bars is from Feeney, Kavanagh \& Regan (2024 ), normalized to the redshift 6 bin and corrected to L46, assuming those with z $>$ 8 are mean luminosity, also assuming that selection effects are independent of redshift.}
\end{figure}

If the present consensus is confirmed that pure density models do not fit the abundance of AGN at redshift 10,
these models will suggest different spectral energy distributions and bolometric corrections or faster buildup
of SMBH. However, fuelling models should also be considered with, perhaps 
, even an extreme duty cycle with QSOs "always on". And, although m8
PBH do not lose mass during galaxy formation, m17--m19 PBH do, and would have been able to supply more fuel at z = 10.

Shen \etal (2020) present a careful analysis of the bolometric LF 
variation with redshift, using
all the data rather than the coarse bins illustrated here. They find both luminosity and density evolution.
The key question for large volume JWST "little red dot" surveys is,
does the naive version of our model, pure density evolution, still fit,
or, are the parameters evolving in some way that sheds light on QSO power?

Models for little red dots supply the SMBH either from accretion on to stellar mass black holes or, as is the case here,
from the properties of dark matter. An alternative in the first category is growth from self interacting dark matter (Jiang \etal 2025) via halo core collapse.
\subsection{A QSO standard candle}
There is an extensive cosmology literature in which quasars are used as standard candles (e.g. Signorini \etal 2024). Figure 2 is supportive of this assumption, although the pure density evolution slope is quite uncertain without higher redshift data.  
   Minimum, mean\footnote{The mean luminosity for equation (6) is 4 $\times$ L46.} and maximum standard 
   luminosities and their dispersion are afforded by the model outlined in $\S$2 as a function of redshift. With higher redshift data arriving from JWST the wealth of data can be expected to amply constrain the simple model presented here.
Traditionally stars and supernovae are seen as standard candles, and galaxies with their complex history less so. Quasars have their roots in PBH which follow a simple law, although accretion may complicate it. This is a physical basis for  QSOs as standard candles.

Risaliti \etal (2023) find a Hubble-Lemaitre diagram with redshifts out to 5 that differs from the $\Lambda$CDM predictions by 4$\sigma$. The standard candle is
from an ultraviolet-xray luminosity relation that is non-linear (logarithmic slope 0.63). There is no theoretical basis for this relation, and our model does not provide one,
as the spectral energy distribution is not modelled here. While equation (4) from fuelling considerations does imply a bolometric luminosity standard candle, and the
parameters, b and f are unlikely to evolve with redshift, the parameter K involves complex astrophysics, and SMBH mass is measureable.

In principle, a maximum luminosity arises from the age of the universe,  the end of the RE, z$_{eq}$, 
 and the origin of SMBH at this time.
That is given by 8GMH$_0 z_{eq}^3~=~c^3$, 
 and M = 2.8 $\times$ 10$^{14}$ M$_\odot$. 
 However, the maximum observed QSO luminosity $\lesssim$ L50, expected for M8.  
 The Eddington luminosity for a 10$^8$ M$_\odot$ black hole (Rees 1984) is 1.3 $\times~10^{46}$, a more likely physical limit (but see King 2024).
\subsection{Radio galaxies}
In hindsight, early results on radio galaxies (Ekers \etal 1981), two-thirds of the sources being double,
the merger, Cen A, and VLBI jets pointed to the presence of black holes and the importance of accretion.
The variety of radio galaxies called at the same early time for unification (Readhead 2015), whereby
diverse properties could be related to the perspective of the observer relative to the axis of the galaxy.

Radio galaxies have a similar LF, 
but scaled down in maximum radio power by 6.35 dex (Pracy \etal 2016). Applying a bolometric correction of 4.5 dex (Marconi \etal 2004 and Panessa \etal 2007), we find correspondence to PBH nuclei of $\sim$10$^{6.5}$, overlapping in nuclear mass with QSOs, and implying a qualitative difference in their accretion state. 
At L46 the QSO LF 
is dlog n / dlog L = --4.7 = --dn/n t/dt, since log~L = 2~log~c +log~m --log~t, and so dn/n t/dt = 4.7.
Continuity requires that dn/dt is constant; so, lifetime in state x $\propto$~n$_x$ where x = quasar, Seyfert, or radio galaxy, if they are all connected by the same galaxy interaction rate. 
If QSO and radio galaxy are different stages of the same phenomenon, the lifetime as radio galaxy is $\sim$10 times that as quasar, according to the LFs.
 The notion of a pause between accretion episodes,
should be added to the unification premise of observer perspective.
Lifetimes of different phases and what can be learned from jets are reviewed by Garofolo \etal 2018).
Duty cycles of radio galaxies are discussed by Shabala \etal (2020).
\section{Conclusions}
The formation of PBH in the radiation dominated universe that ended at z $\approx$ 3400 
has been considered here to supply a generation of SMBHs, whose average luminosity is 4~$\times$~10$^{46}$ ergs/s, that is 10$^{13}$ L$_\odot$. There is evidence that PBH trace
their origin as density enhancements to seeds generated in inflation that were then built on during the RE. Their DNA, so to speak, bears an imprint of inflation in the IMF, as Figure 1 bears out. 
A PBH mass function of n $\sim$ m$^{-1}$
is indicated by the luminosity function fit. As a constituent of halos, PBH also supply some of the fuel for SMBH accretion.
The predicted luminosity function is a very good fit to what is observed and
is normalised at only one point.  

Although the caveat was noted by the first proponents of PBH as QSO progenitors, the possibility that  formation of 10$^7$ M$_\odot$ PBH might leave a mark on Big Bang Nucleosynthesis has not been considered here  
($cf$ Sanderbeck, Bird, \& Haiman, 2021).
Given readily available light element abundance data, plus the large infall radii of SMBH we outlined in $\S$2, the consequences should be explored in detail with calculations of relativistic infall and the resulting density enhancements. These infall regions are galaxy sized. A dispersion in galaxy initial helium abundances, for example, might be expected.

The luminosity density of these QSOs  is a function of the efficiency of black hole fuel consumption, the mean mass of both the  PBH that supply the dark matter for the consumed galaxies and the SMBH, the velocity of SMBH-galaxy collisions, the mean redshift of formation of galaxies and the enhancement by gravity of a collision rate expected from a geometric cross section.
The parameters characterising these processes are degenerate in their effect on QSO numbers, but some will be able to be simulated and then removed as parameters from the model. 
 This will be strengthened by further detailed work on the astrophysical processes involved.
In the near future number densities at higher redshifts
from JWST, not limited by the small number statistics in Figure 2, will extend the observational testing of fuel burning models like this.
Detection of CMB distortions or a primordial helium abundance
dispersion might also signal the presence of SMBH at early times.

To summarize:
\begin{itemize}
	\item If supermassive PBH are formed towards the end of the radiation
		dominated era, they can provide the nuclei of active galaxies
		at redshifts as high and higher than those now observed with JWST.
	\item If the IMF of PBHs is such that there are equal numbers in
		each decade of mass, a QSO LF is expected which is in
		agreement with observations of their bolometric luminosity and number density.

	\item The proviso is that the product of the following quantities
		has an upper limit: the fraction of the dark matter that is subsolar PBHs, the efficiency of the SMBH turning accreted mass into energy and 
		the collisional velocity of galaxies supporting the accretion.

	\item This model supports unified theories of active galaxies, and would support the existence of IMBHs in galaxies, if they are confirmed.
\end{itemize}

\subsection*{References}
\noindent
Antonucci, R. 2013, Nature, 495, 165\\
Alcock, C. \etal 1998, ApJL, 499, L9\\
Belotsky, K. \etal 2019, 
Eur. Phys. J. C 79: 246\\
Bicknell, G. \& Henriksen 1979, MNRAS, 232, 670\\
Blandford, R. \& Begelman, M. 1999, MNRAS, 303, L1\\
Cai, Y.-F. \etal 2024, 
 Mechanics \& Astronomy 67, 259512 \\
Carr, B. \& Kuhnel, F. 2021a, arxiv 21100282\\
Carr, B., Kohri, K., Sendouda, Y. \& Yokoyama, J. 2021b, RPP, 84, id.116902\\
Clayton, D. 1968, {\it Principles of Stellar Evolution \& Nucleosynthesis},
McGraw-Hill, New~York\\
Conselice, C. \etal 2020, ApJ, 890, 8\\
Davies, M., Miller, M. \& Bellovary, J.2011, ApJL 740,  L42\\
Davis, B. \etal 2024, ApJ, 971, 123\\
Dayal, P. \& Maiolino, R. 2025, arxiv 2506.08116\\
Deng, H. 2021, JCAP, 11, 054\\
Ekers, R. \etal 1981, A\&A, 101, 194\\
Fan, X. 2010, AIP Conf Proc, 1279, 44\\
Feeney, J., Kavanagh, P. \& Regan, J. 2024, arxiv 2409-13441\\
Fixsen, D. \etal 1996, ApJ, 473, 576\\ 
Frampton, P. 2018, Mod Phys Lett A, 31, 1850176\\
Garofolo, D., Singh, C. \& Zack, A. 2018, Nat Sci Rep 8, 15097\\
Greene, J., Strader, J. \& Ho, L. 2020, ARAA, 58, 257\\
Harada, T. 2024, Univ, 10, 444\\
Hooper, D., Ireland, A. \& Krnjaic, G. 2024, JCAP, 4, 21\\
Jiang, F. \etal 2025, arxiv 2503.23710\\
King, A. 2024, MNRAS, 536, L1\\
Kocevski, D. \etal 2025, ApJ, 986, 126\\
Kovacs, Z., Gergely, L. \& Biermann, P. 2011, MNRAS, 416, 991\\
K\"uhnel, F., Rampf, C. \& Sandstad, M. 2016, PRD, 94, 3504\\
Labb\'e, I. \etal 2025, ApJ, 978, L92\\
Li, Y.-Y., Huang, H.-L. \& Piao, Y.-S. 2025, arxiv 2509.03152\\
Liu, B. \& Bromm, V. 2025, ApJL, 937, L30\\
Lupi, A. M. \etal 2014, MNRAS, 442, 3616\\
Lynden-Bell, D \& Rees, M. 1971, MNRAS, 152, 461\\
Mandal, A., Woo, J.-K. \& Wang, S. 2025, arxiv 2502.19184\\ 
Matthee, J. \etal 2024, ApJ, 963, 129\\ 
Marconi, A. \etal 2004, MNRAS, 351, 169\\
Moniez, M. 2001, {\it Cosmological physics with gravitational lensing}: Proc.  XXXVth Rencontres de Moriond, Les Arcs, France. Eds: J. Thanh Van Tran, Y. Mellier, \& M. Moniez. Les Ulis: EDP Sciences, 2001, p.3\\
Moniez, M. \etal 2017, A\&A, 604, 124\\
Mould, J.  2025, ApJ, 984, 59\\
Musco, I. 2019, PRD, 100, 123524\\
Nakama, T., Carr, B. \& Silk, J. 2018, PRD, 97, 043525\\
\"Ozsoy, O. \& Tasinato, G. 2023, Univ, 9, 203\\
Porras-Valverde, A. \etal 2025, arxiv 2504.11566\\
Panessa, F. \etal 2006, A\&A, 455, 73\\
Peebles, J. 1976, Ap\&SS, 45, 3\\
Planck collaboration, 2020, A\&A, 641, 9\\
Pracy, M. \etal 2016, MNRAS, 460, 2\\
Prole, L. \& Regan, J. 2025, ApJ, 987, 185\\
Qiang, D.-C. and Wei, H. 2020, JCAP, 04, 023\\
Readhead, A. 2015, Proc. TORUS2015, editors: P. Gandhi \& S. Hoenig\\
Rees, M. 1984, ARAA, 22, 471\\
Ren, K. \& Trenti, M. 2021, ApJ, 923, 110\\
Risaliti, G. \etal 2023, AN, 344, 20230054\\
Sanders, D., Phinney, E. S., Neugebauer, G., Soifer, B. T., Matthews, K. 1989, ApJ, 347, 29\\
Sanderbeck, P., Bird, S. \& Haiman, Z. 2021, arxiv 2109.05035\\
Schechter, P. 1976, ApJ, 203, 297\\
Schmidt, M 1963, Nature, 197, 1040\\
Schmidt, M. 1968, ApJ, 151, 393\\
Shabala, S. \etal 2020, MNRAS, 496, 1706\\
Shakura, N. \& Sunyaev, R. 1973, A\&A, 24, 337\\
Shen, X. \etal 2020, MNRAS, 495, 3252\\
Signorini, M. \etal 2024, A\&A, 687, 32\\
Sobrihno, J. \& Augusto, P. 2024, MNRAS, 531, L40\\
Soszynski, I. 2024, CoSka, 54, 234\\
Volonteri, M., Habouzit, M. \& Colpi, M. 2021, Nature Reviews Physics 3,  732\\
Wang, Z.-He, Huang, H.-L. \& Piao, Y.-S. 2025, , PRD, 111, 83553\\
Winkel, N. \etal 2025, ApJ, 978, 115\\
Yang, J. \etal 2025, PRD, 111, id.043522\\
Zegeye, D., Inomata, K. \& Hu, W. 2022, PRD 105, 103535\\
Zhang, H. \etal 2024, MNRAS, 529, 2777\\
Zhang, S., Liu, B., Bromm, V. \&  K\"uhnel, F. 2025, arxiv 2512.14066\\
Ziparo, F., Gallerani, S. \& Ferrara, A. 2025, JCAP, 04, 40\\
\subsection*{Data Availability}
There are no data published in this paper to make available.
\subsection*{Acknowledgements}
We would like to thank our colleagues in the Swinburne CAS \& ANU RSAA microlensing team and the ARC Centre of Excellence for Dark Matter Particle Physics for stimulating discussions, and we acknowledge ARC grant CE200100008 which, together with the five Australian university nodes, funds the centre's research. We also acknowledge the generous intellectual support of the centre's international partners. We thank Matthew Bailes for reminding us about the IGM and Elaine Sadler for presenting the radio galaxy luminosity function at the recent centenary of Mt.  Stromlo Observatory  
and Karl Glazebrook for an update on little red dots. We are grateful to the referees for improving the paper. 

Jeremy Mould adds his thanks to 
 Alvio Renzini who developed the fuel burning theorem in stellar evolution. In memory of our late colleagues at Caltech, Jesse Greenstein, Wal Sargent and Martin Schmidt, quasar pundits all.
\end{document}